# The Stochastic Field Transport associated with the Slab ITG Modes

J W Connor[1,2,3], R J Hastie[1] and A Zocco[1,3]

[1] EURATOM/CCFE Fusion Association, Culham Science Centre, Abingdon, Oxon, OX14 3DB, UK
[2] Imperial College of Science, Technology and Medicine, London SW7 2BZ, UK
[3] Rudolf Peierls Centre for Theoretical Physics, 1 Keble Road, Oxford, OX1 3NP, UK

**Abstract**

Many models for anomalous transport consider the turbulent $E \times B$ transport arising from electrostatic micro-instabilities. In this paper we investigate whether the perturbed magnetic field that is associated with such instabilities at small but finite values of β can lead to significant stochastic magnetic field transport. Using the tearing parity, long wave-length ion temperature gradient (ITG) modes in a plasma slab with magnetic shear as an example, we calculate the amplitude of the perturbed magnetic field that results at the resonant surface for the case when the plasma dissipation is given by the semi-collisional electron model. The resulting stochastic field transport is estimated and also compared with an estimate for the $E \times B$ transport due to the ITG mode.

.

## 1. Introduction

Calculations of anomalous transport due to micro-instabilities tend to concentrate on the contribution from the perturbed $E \times B$ drifts arising from electrostatic perturbations [1], although there are a number of studies of the cross-field transport arising from the stochastic magnetic fields associated with some electromagnetic instabilities, e.g. micro-tearing modes [2, 3, 4, 5]. There remains the possibility that the electromagnetic perturbation that inevitably accompanies a predominately electrostatic instability, such as the ion temperature gradient (ITG) mode, due to a small but finite value of $\beta = 2\mu_0 nT/B^2$, i.e. $\beta << 1$, can produce significant stochastic field transport, possibly exceeding the original $E \times B$ contribution. The purpose of the present paper is to investigate this possibility using the long wave-length, slab-like, tearing parity ITG instabilities as a simple example [6]. This complements a recent study of stochastic magnetic transport arising from numerical simulations of ITG turbulence [7] that found non-linearly excited stable micro-tearing modes, rather than the unstable tearing parity ITG modes themselves, were responsible for the stochastic magnetic field transport.

The electrostatic ITG perturbation in slab geometry is dominated by long wave-lengths, with $k_\perp \rho_i \leq 1$ where $\rho_i = (2m_i T_i)^{1/2}/eB$ is the ion Larmor radius and $k_\perp$ is the wave-number perpendicular to the magnetic field. However any magnetic field reconnection due to an associated electromagnetic component, as is required to produce a stochastic magnetic field, will occur at the much smaller scales where electron dissipative effects enter. Thus in modelling the ion dynamics we must consider a full Larmor radius (FLR) response for the ions to connect the long wavelength ITG perturbation to the reconnecting region of $k_\perp$. As an experimentally relevant model for the electrons we consider a semi-collisional description, when $\omega \nu_e \sim k_\parallel^2 v_{the}^2$, where $\omega$ is the mode frequency, $\nu_e$ is the electron collision frequency, $v_{the}$ is the electron thermal speed, $v_{the} = (2T_e / m_e)^{1/2}$, and $k_\parallel = k_y x / L_s$ is the wave-length along the magnetic field. Here $L_s = Rq^2 / (rdq/dr)$ is the magnetic shear length and $k_y$ is the wave-number perpendicular to the magnetic field, **B**, but within the equilibrium magnetic surface. Such semi-collisional effects are significant when electron collisional transport processes along the magnetic field compete with the mode frequency, allowing us to define a 'semi-collisional width',
$\delta = e^{-i\pi/4} \left( \omega \nu_{ei} / k_y^2 (T_e / m_e) \right)^{1/2} L_s$.

To simplify the problem further we consider the 'flat-density' limit where the density scale-length, $L_n \to \infty$, order the wavelengths so that $b_0 = k_y^2 \rho_i^2 / 2 \sim (L_{Ti} / L_s)^{1/2} \ll 1$ and normalise the frequencies such that $\omega = (\omega_{*Ti} T_i / m_i L_s^2)^{1/3} \bar{\omega}$. Here we have introduced the diamagnetic-like frequency $\omega_{*Ti} = k_y T_i / eBL_{Ti}$ with $L_{Ti}$ the scale-length of the ion temperature gradient. To develop a systematic ordering scheme, we consider $L_{Ti} / L_s \ll 1$, i.e. sharp ion temperature gradients, and introduce a parameter $\alpha = (b_0^2 L_s / L_{Ti})^{2/3} \sim 0(1)$ so that $b_0 \sim (L_{Ti} / L_s)^{1/2}$.

In standard tearing mode theory the amount of reconnection at a resonant surface depends in part on a quantity $\Delta'$ which represents the instability drive due to the current gradient. An equivalent quantity, $\Delta^*$, encapsulating the effects of the tearing parity, long wave-length electrostatic ITG mode, appears in the present theory as a boundary condition at small values of k in Fourier space (although at larger values than those corresponding to the ITG mode itself, i.e. at $k_x \rho_i \gg b_0^{1/2}$) and plays a part in determining the electromagnetic response in the vicinity of the resonant surface, where $k_x$ is large [8]. For the case of the semi-collisional electron model, this reconnecting region corresponds to a radial wave-number given by $k_x \sim 1/\delta_0$, where we have introduced a real parameter, $\delta_0$, to characterise the semi-collisional layer width. This is defined as $\delta_0 = |\delta(\omega)|$, evaluated at $\omega = (\omega_{*Ti} T_i / m_i L_s^2)^{1/3}$. Thus $\delta_0 / \rho_i = b_0^{-5/12} (m_e / 2m_i)^{1/4} (T_i / T_e)^{1/2} (\nu_{ei} L_s / v_{the})^{1/2} (L_s / L_{Ti})^{1/6}$. The calculation below aims to quantify the amount of reconnection and determine its implications for stochastic field transport.

In Section 2 we provide the fundamental equations describing the electromagnetic ITG modes and sketch the structure of the calculation. Section 3 determines the characteristics of the long wave-length electrostatic ITG modes while Section 4 calculates the magnetic perturbations they drive in the same wave-length region. In order to connect this to the electromagnetic response in the electron layer, in Section 5 we obtain the corresponding solution in the intermediate ion FLR region where $k_x\rho_i \sim O(1)$, matching it to the long wave-length solution using the quantity $\Delta^*$ introduced above. In Section 6 we calculate the solution in the electron layer and, by matching it to the intermediate ion region solution in Section 7, determine the magnetic perturbation at the resonant surface in Section 8. In Section 9 we estimate the resulting stochastic magnetic field transport and compare it with the electrostatic transport associated with the original ITG mode $E \times B$. Conclusions are drawn in Section 10.

## 2. Fundamental Equations

In this section we establish the quasi-neutrality equation and Ampère's equation which determine the electromagnetic response to long wavelength, slab-like, electrostatic ITG modes and outline the sequence of steps in the subsequent calculation.

The equations describing the finite Larmor radius response of the ions to an electromagnetic perturbation in sheared slab geometry are derived in Appendix A. Finite ion Larmor radius effects effects were discussed in Refs. [8, 9], but it is necessary to calculate the 'ion sound' corrections in order to describe the slab ITG mode. Likewise the equations describing the electrons for semi-collisional plasma, i.e. when collisional transport processes along the magnetic field compete with the mode frequency, are derived in Appendix B. Again such a model was also discussed in Refs. [8, 9], but it is now necessary to account for ion parallel motion in the Ohm's law. Substituting the calculated perturbed ion and electron densities into the quasi-neutrality equation provides one of the governing equations, while inserting the calculated perturbed current into Ampère's law provides the other.

To obtain the quasi-neutrality equation we use eqn. (A.10), which includes the 'ion sound' corrections of $0(k_\parallel^2 v_\parallel^2/\omega^2)$, for the perturbed ion density, $\delta n_i$, while the perturbed electron density, $\delta n_e$, is provided by the electron continuity equation, eqn. (B.1). However the latter equation involves the perturbed electron parallel velocity, $u_{\parallel e}$, which is obtained in terms of the perturbed parallel component of the vector potential, $A_\parallel$, and the perturbed parallel ion velocity, $u_{\parallel i}$, from eqn. (A.11) with the aid of Ampère's equation, eqn. (A.9). Similarly we use eqn. (B.6) for the perturbed parallel current, $j_\parallel$, again substituting for $u_{\parallel i}$ from eqn. (A.11), to obtain Ampère's law for $A_\parallel$.

Introducing the flat-density ordering discussed in the Introduction, namely $b_0 = k_y^2\rho_i^2/2 \sim \left(L_T/L_s\right)^{1/2}$, with $L_{Ti}/L_s \ll 1$, and normalising the frequencies such that $\omega = \left(\eta_i\omega_{*i}\left(T_i/m_iL_s^2\right)\right)^{1/3}\overline{\omega}$, simplifies the equations considerably. The quasi-neutrality equation takes the form:

$$\frac{i\tau\alpha^2}{\overline{\omega}^2\hat{\beta}_T b_0^2}\frac{d}{d\overline{k}}\left(\overline{k}_\perp^2 \hat{A}\right) = H\hat{\varphi} - \frac{b_0}{\overline{\omega}^3}\left[\frac{d}{d\overline{k}}\left(\frac{d\hat{\alpha}_1}{d\overline{k}}\hat{\varphi}\right) - i\frac{d\hat{\alpha}_1}{d\overline{k}}\hat{A}\right]. \tag{1}$$

Ampère's law can be written $b_0$

$$\frac{\tau^2\alpha^2}{\overline{\omega}^2\hat{\beta}_T b_0^2}\frac{\delta^2}{\rho_i^2}\left(\overline{k}_\perp^2 \hat{A}\right) = \frac{\left(\hat{\sigma}_0 + d_1(x/\delta)^2\right)}{\left(1 + d_0(x/\delta)^2 + d_1(x/\delta)^4\right)}\left(i\frac{d}{d\overline{k}}\hat{\varphi} + \hat{A}\right)$$
$$+ \frac{2\tau b_0}{\overline{\omega}^3}\frac{x^2}{\rho_i^2}\frac{\left(d_2 + d_1(x/\delta)^2\right)}{\left(1 + d_0(x/\delta)^2 + d_1(x/\delta)^4\right)}\left[\hat{\alpha}_1\left(i\frac{d}{d\overline{k}}\hat{\varphi} + \hat{A}\right) + \frac{i}{2}\frac{d\hat{\alpha}_1}{d\overline{k}}\hat{\varphi}\right], \tag{2}$$

where we have introduced $\alpha = \left(b_0^2 L_s/L_T\right)^{2/3} \sim 0(1)$ and $\overline{k} = k_x\rho_i$. In eqn. (2) we have used a mixed notation in x- and $k_x$- space, so that, if we consider these equations in $k_x$-space for example, we identify $x = -id/dk_x$ and *vice versa*. Other notation is as follows:

$$\hat{\varphi} = e\varphi/T_i, \quad \hat{A} = \left(\omega L_s/\overline{k}_y\right)eA_\parallel/T_i, \quad \hat{\beta}_T = \beta_e\left(L_s/L_{Ti}\right)^2/2 \text{ with } \beta_e = 2\mu_0 nT_e/B^2,$$

$$\tau = T_e/T_i, \quad H = \Gamma_0(b) - 1 + \frac{\alpha b}{b_0\overline{\omega}}[\Gamma_0(b) - \Gamma_1(b)] - \frac{b_0}{\overline{\omega}^3 b}\left[\left(\overline{k}^2 - \overline{k}_y^2\right)\frac{d\hat{\alpha}_1}{d\overline{k}^2} + \overline{k}^2\hat{\alpha}_2\right],$$

$$\hat{\alpha}_1 = \Gamma_0(b) - b(\Gamma_0 - \Gamma_1), \quad \hat{\alpha}_2 = \left[2(b-1)^2\Gamma_0(b) - b(2b-3)\Gamma_1(b)\right],$$

$$\hat{\sigma}_0 = 1.71\frac{\tau\alpha}{\overline{\omega}b_0}\frac{L_{Ti}}{L_{Te}}, \quad d_0 = 5.1, \quad d_1 = 2.1, \quad d_2 = 2.9,$$

$$b = \frac{\overline{k}_\perp^2}{2} = \frac{k_\perp^2\rho_i^2}{2}, \text{ where } k_\perp^2 = k_y^2 + k_x^2 \quad b_0 = \frac{k_y^2\rho_i^2}{2}, \quad \delta = \exp(-i\pi/4)\left(\frac{m_e\omega\nu_{ei}}{k_y^2 T_e}\right)^{1/2}L_s. \tag{3}$$

These equations can be used to determine the electromagnetic perturbations at the resonant surface driven by the long wave-length electrostatic ITG modes and estimate the associated stochastic field transport. The approach involves the following sequence of calculations based on an ordering of terms in $\hat{\beta}_T b_0^2 \ll 1$, but with $\hat{\beta}_T \sim 0(1)$:

- By including the ion sound terms, determine the characteristics of the 'local', long wave-length (i.e. $k_x \sim k_y \sim b_0^{1/2}$), tearing parity, electrostatic ITG modes: $\varphi_{ITG}$.
- Calculate the associated, long wave-length, electromagnetic correction, $A_\parallel^{(1)}$, of $0(\hat{\beta}_T b_0^2)$, which introduces a constant of integration, $c_0$, corresponding to a vacuum solution of Ampère's equation.

- Calculate $\varphi^{(1)}$, the correction to $\varphi_{ITG}$ driven by $A_\parallel^{(1)}$ in the long wave-length region. This provides a low $k_x$ boundary condition for determining the 'non-local' $\varphi^{(1)}$ in the ion FLR region ( $k_x \sim \rho_i^{-1} \gg k_y$ ).
- Calculate $\varphi^{(1)}$ for $k_x \sim \rho_i^{-1} \gg k_y$ using the above boundary condition (we also calculate the second order correction in $\hat{\beta}_T b_0^2$, $\varphi^{(2)}$, which helps to clarify the subsequent matching to the semi-collisional electron region).
- Determine the contributions to $A_\parallel$, driven by $\varphi_{ITG}$, $\varphi^{(1)}$ and $\varphi^{(2)}$ in the ion FLR region.
- Calculate $A_\parallel$ in the semi-collisional electron region ( $k_x \sim \delta_0^{-1}$ ) satisfying the appropriate (i.e. even) boundary condition at $x = 0$.
- Determine the constant $c_0$ by matching electron and ion region solutions for $A_\parallel$; this matching also determines $A_\parallel(0)$, the amount of reconnection at the resonant surface, $x = 0$, in terms of $\varphi_{ITG}$
- Calculate $\delta B_x$, the 'radial' perturbed magnetic field at the resonant surface, from $A_\parallel(0)$ and estimate the resulting stochastic field transport using the Rechester - Rosenbluth formula [10]

The above expansion involves a separation of electron and ion scales: $\delta_0 \ll \rho_i$. For consistency, this condition imposes a constraint, $1 \gg b_0 \gg (m_e/m_i)^{3/5} (\nu_e L_s/v_{the})^{6/5} (L_s/L_{Ti})^{2/5}$, on our parameters; this is readily satisfied in hot tokamaks.

## 3. Slab ITG Mode: $k \quad k_y$

Before discussing the slab ITG mode in detail it is helpful to consider eqn. (2) in the 'ion region', $x \sim \rho_i \gg \delta_0$ (i.e. $k \sim \rho_i^{-1} \ll \delta_0^{-1}$), where it simplifies considerably:

$$\frac{\tau^2 \alpha^2}{\overline{\omega}^2 \hat{\beta}_T b_0^2} \frac{d^2}{d\overline{k}^2}\left(\overline{k}_\perp^2 \hat{A}\right) = -\left(i\frac{d}{d\overline{k}}\hat{\varphi} + \hat{A}\right) + \frac{2\tau b_0}{\overline{\omega}^3} \frac{d^2}{d\overline{k}^2}\left[\hat{\alpha}_1 \left(i\frac{d}{d\overline{k}}\hat{\varphi} + \hat{A}\right) + \frac{i}{2}\frac{d\hat{\alpha}_1}{d\overline{k}}\hat{\varphi}\right]. \quad (4)$$

To describe the electrostatic ITG mode we consider the limit $\hat{\beta}_T b_0^2 \ll 1$ so that we can ignore $\hat{A}$ on the right-hand side of eqn. (4). Taking the long wavelength limit ($b \ll 1$), when $\hat{\alpha} = 1 + 0(\overline{k}^2)$, we can integrate once to yield

$$\frac{i\alpha\tau}{\overline{\omega}\hat{\beta}_T b_0} \frac{d}{d\overline{k}}\left(\overline{k}_\perp^2 \hat{A}\right) = \frac{\overline{\omega} b_0}{\alpha\tau}\left(1 - \frac{2\tau b_0}{\overline{\omega}^3}\frac{d^2}{d\overline{k}^2}\right)\hat{\varphi} \quad . \quad (5)$$

Substituting for $\hat{A}$ from eqn. (1), where $H \simeq \alpha b/\overline{\omega} b_0$ and $\hat{\alpha}_1 \simeq 1$ in the same limit, so that we can ignore the second term on the right-hand side, we obtain the equation for the long wave-length, electrostatic ITG modes:

$$L^{(0)}\hat{\varphi}^{(0)} \equiv \frac{2\tau b_0}{\overline{\omega}^3}\frac{d^2}{d\overline{k}^2}\hat{\varphi}^{(0)} + \left(\frac{\alpha\tau}{2b_0\overline{\omega}}\overline{k}_\perp^2 - 1\right)\hat{\varphi}^{(0)} = 0 \ . \tag{6}$$

Since we are interested in electromagnetic effects with finite $A_\parallel$ at the resonant surface we consider the tearing parity solutions, i.e. those odd in $\overline{k}$:

$$\hat{\varphi}_n^{(0)} \equiv \hat{\varphi}_{ITG} = c_n H_{2n+1}\left((\lambda_n/b_0)^{1/2}\overline{k}\right)\exp(-\lambda_n\overline{k}^2/2b_0)\varphi_0 \ . \tag{7}$$

Here the index n labels the various harmonic solutions of eqn. (6) with eigenvalues $\overline{\omega} = \overline{\omega}_n$, $c_n = \left(2^{2n+1}\sqrt{\pi}\,\Gamma(2n+2)\right)^{-1/2}$ is a normalisation constant, $H_{2n+1}$ are the Hermite polynomials and $\lambda_n = -i\alpha^{1/2}\overline{\omega}_n/2$, so that $\mathrm{Re}\,\lambda_n > 0$ when $\mathrm{Im}\,\overline{\omega}_n > 0$. We have labelled the amplitude of the ITG mode by $\varphi_0$ and intend, ultimately, to relate the amplitude of the perturbed magnetic field to it.

The condition on the eigenvalue, $\overline{\omega}_n$, is

$$\overline{\omega}_n^2 - \alpha\tau\overline{\omega}_n - (2n+1)i\alpha^{1/2}\tau = 0 \ . \tag{8}$$

The solution for $\overline{\omega}_n$ depends on the parameter $\alpha$. It will be interesting later to consider two limiting cases:

$$\text{(i)} \quad \alpha \ll \left[4(2n+1)/\tau\right]^{2/3}, \tag{9}$$

when $\overline{\omega}_n = \left[(2n+1)\tau\right]^{1/2}\alpha^{1/4}e^{i\pi/4}$ and $\lambda_n = \left[(2n+1)\tau\right]^{1/2}\alpha^{3/4}e^{-i\pi/4}/2$;

$$\text{(ii)} \quad \alpha \gg \left[4(2n+1)/\tau\right]^{2/3}, \tag{10}$$

when $\overline{\omega}_n = \tau\alpha + (2n+1)i/\alpha^{1/2}$ and $\lambda_n = \left[-i\tau\alpha^{3/2} + (2n+1)\right]/2$. Clearly for higher values of n the result (i) prevails for all $\alpha$. Furthermore, the higher n harmonics, which correspond to higher radial harmonics in x-space, have larger growth rates and radial wave-numbers. Thus forming the expectation value of $\overline{k}^2$ over the eigen-function (7), we obtain $\langle\overline{k}^2\rangle_n = (n+1/2)b_0/\lambda_n \sim (n+1/2)^{1/2}b_0$. Of course the maximum acceptable value of n is limited by the two approximations: (a) $\delta_0/\rho_i \ll 1$, where $\delta_0 \propto \overline{\omega}_n^{1/2} \propto n^{1/4}$; (b) $\langle\overline{k}^2\rangle_{ITG} \ll 1$, with $\langle\overline{k}^2\rangle_{ITG} \sim nb_0/\lambda_n \propto n^{1/2}$.

## 4. Electromagnetic Corrections for $k_x \sim k_y$

This section provides a calculation of the long wave-length, electromagnetic perturbations driven by the ITG modes. Thus, in next order in $\hat{\beta}_T b_0^2$ we obtain $\hat{A}^{(1)}$ by integrating eqn. (5):

$$\bar{k}_\perp^2 \hat{A}^{(1)} = c_0 - i\frac{\bar{\omega}_n \hat{\beta}_T b_0}{2\tau\alpha} \int_0^{\bar{k}} d\bar{k}\, \bar{k}_\perp^2 \hat{\varphi}_{ITG} , \qquad (11)$$

with $c_0$ a constant of integration of order $\hat{\beta}_T b_0^2$. In turn $\hat{A}^{(1)}$ drives a correction to $\hat{\varphi}_{ITG}$, $\hat{\varphi}^{(1)}$, which is non-local in $\bar{k}$, i.e. not restricted to $k_x \sim k_y$. However, we first determine $\hat{\varphi}^{(1)}$ for $k_x \sim k_y$ to provide the boundary condition for $\hat{\varphi}^{(1)}$ in the region $\bar{k} \sim 1$, which we calculate in the next section.

Integrating eqn. (4) once in $\bar{k}$ and substituting for $\bar{k}_\perp^2 \hat{A}^{(1)}$ from eqn. (5), we have

$$L^{(0)}\hat{\varphi}^{(1)} = c_1 - i\int_0^{\bar{k}} d\bar{k}\, \hat{A}^{(1)} + i\frac{2\tau b_0}{\bar{\omega}_n^3}\frac{d}{d\bar{k}}\hat{A}^{(1)} - \delta\bar{\omega}_n \left.\frac{\partial L^{(0)}}{\partial \bar{\omega}}\right|_{\bar{\omega}_n} \hat{\varphi}_{ITG} \equiv R(\bar{k}), \qquad (12)$$

where $c_1$ is a second constant of integration and $\delta\bar{\omega}_n$ is a correction to $\bar{\omega}_n$. Since we seek a tearing parity solution $\hat{\varphi}^{(1)}$ (i.e. odd in $\bar{k}$), we take $c_1 = 0$. On the other hand, the constant $c_0$ in eqn. (11) emerges, ultimately, as the 'eigenvalue' of the present calculation: i.e. it will eventually be determined by matching solutions through an intermediate region where $\bar{k} \sim 0(1)$, to an electron region (a layer around $x = 0$ in configuration space, corresponding to the region $\bar{k} \geq \rho_i/\delta_0$ in transform space) and application of appropriate boundary conditions as $\bar{k} \to \infty$. Along with the final determination of $c_0$ the complete structure of the electromagnetic eigenmode, i.e. both $\hat{A}(\bar{k})$ and $\hat{\varphi}(\bar{k})$, will then be specified. In particular the amplitude of the magnetic perturbation at the resonant surface, $\delta B_x/B$ at $x = 0$, relative to that of the underlying 'electrostatic' ITG mode, which remains arbitrary within this linear theory.

Annihilating $\hat{\varphi}^{(1)}$ in eqn. (12) by applying the operation, $\int_0^\infty d\bar{k}\, \hat{\varphi}_{ITG}(\bar{k})...$ we obtain the condition

$$\int_0^\infty d\bar{k}\, \hat{\varphi}_{ITG}(\bar{k}) R(\bar{k}) = 0 , \qquad (13)$$

which provides an equation for $\delta\bar{\omega}_n$, although this is not needed for our present purposes.

A solution for $\hat{\varphi}^{(1)}$ in terms of $R(\bar{k})$ from eqn. (12) satisfying the appropriate boundary conditions can be obtained by the method of variation of parameters using the two solutions of the homogeneous eqn. (6). These are the parabolic cylinder functions: $U(-(n+1/2), z)$ and $V(-(n+1/2), z)$ where $z = (2\lambda/b_0)^{1/2}\bar{k}$ [11]. [Note solution (7) is essentially the exponentially decaying function $U(-(n+1/2), z)$ and condition (13) ensures there is no exponential growth from the $V(-(n+1/2), z)$ contribution.]

However, we can determine the required features of the solution without formally constructing this expression. Thus at small k we require $\hat{\varphi}^{(1)}$ to satisfy the ideal MHD condition, $E_\parallel = 0$. It is evident from eqn. (12) that the dominant terms when $\bar{k}^2 << b_0$ originate from the 'ion-sound' terms and are

$$\frac{\tau b_0}{\bar{\omega}_0^3} \frac{d}{d\bar{k}}\left(i\frac{d}{d\bar{k}}\hat{\varphi}^{(1)} + \hat{A}^{(1)}\right) \propto \frac{d}{d\bar{k}}\hat{E}_\parallel. \tag{14}$$

Clearly in x-space this implies $xE_\parallel \simeq 0$, i.e. $E_\parallel \simeq 0$ at large x. In $\bar{k}$-space, integration of eqn. (14) introduces an arbitrary constant but this becomes a $\delta$-function in x-space and is therefore irrelevant to the large x region. The small $\bar{k}$ form of the solution in terms of parabolic cylinder functions discussed above confirms this result.

The information we actually need from $\hat{\varphi}^{(1)}$ is its behaviour when $1 > \bar{k} > b_0^{1/2}$ in order to provide a low $\bar{k}$ boundary condition for the solution in the $\bar{k} \sim 1$ region. This is readily obtained from eqn. (12), recalling the definition of the operator L in eqn. (6), to yield:

$$\hat{\varphi}^{(1)} \simeq \frac{2b_0\bar{\omega}_n}{\alpha\tau}\frac{R(\bar{k})}{\bar{k}^2}, \tag{15}$$

a result again confirmed from the large z asymptotic form of the solution in terms of parabolic cylinder functions. It remains to determine the form of $R(\bar{k})$ in this region of $\bar{k}$. Clearly from its definition in eqn. (12), $R(\bar{k})$ is dominated by the contribution from the integral term $\int_0^{\bar{k}} d\bar{k}\,\hat{A}^{(1)}$, where $\hat{A}^{(1)}$ is given by eqn. (11), since the term in $\delta\bar{\omega}$ involves the localised solution $\hat{\varphi}_{ITG}$ obtained in eqn. (7) and the contribution from $d\hat{A}^{(1)}/d\bar{k}$ is smaller by a factor $b_0/\bar{k}^2$. Evaluating this dominant term we obtain

$$R(\bar{k}) = -i\frac{c_0}{k_y}\tan^{-1}\left(\frac{\bar{k}}{k_y}\right) - \frac{\bar{\omega}_n\hat{\beta}_T b_0 \varphi_0}{2\alpha\tau}\int_0^{\bar{k}}\frac{d\bar{k}}{\bar{k}_\perp^2}\int_0^{\bar{k}}d\bar{k}' c_n H_{2n+1}\left(\left(\frac{\lambda_n}{b_0}\right)^{1/2}\bar{k}'\right)\bar{k}'^2\exp\left(-\frac{\lambda_n\bar{k}'^2}{2b_0}\right).$$

(16)

Integrating by parts yields:

$$R(\bar{k}) = -i\frac{c_0}{k_y}\tan^{-1}\left(\frac{\bar{k}}{k_y}\right) - \frac{\bar{\omega}_n\hat{\beta}_T b_0^2 \varphi_0}{2\alpha\tau}$$
$$\times c_n\left(\frac{2}{\lambda_n}\right)^{1/2}\int_0^y dy'\left[\tan^{-1}\left(\frac{y}{\sqrt{2\lambda_n}}\right) - \tan^{-1}\left(\frac{y'}{\sqrt{2\lambda_n}}\right)\right]\left(1 + \frac{y'^2}{2\lambda_n}\right)\exp\left(-\frac{y'^2}{2}\right)H_{2n+1}(y'),$$

(17)

where $y' = \exp(-i\pi/4)\left(\bar{\omega}_n\alpha^{1/2}/2b_0\right)^{1/2}\bar{k}'$. Evaluating expression (17) in the limit $1 \gg \bar{k} \gg \bar{k}_y = (2b_0)^{1/2}$ we obtain

$$R(\bar{k}) \sim R_0 + \frac{R_1}{\bar{k}}$$
$$\equiv -i\frac{\pi}{2}\frac{1}{\sqrt{2b_0}}\left\{c_0 - i\frac{\bar{\omega}_n\hat{\beta}_T b_0^{5/2}}{\alpha\tau}F_n(\lambda_n)\varphi_0\right\} + i\left\{c_0 - i\frac{\bar{\omega}_n\hat{\beta}_T b_0^{5/2}}{\alpha\tau}G_n(\lambda_n)\varphi_0\right\}\frac{1}{\bar{k}},$$

(18)

where

$$F_n(\lambda_n) = \frac{c_n}{\sqrt{\lambda_n}}\int_0^\infty dy'\left[1 - \frac{2}{\pi}\tan^{-1}\left(\frac{y'}{\sqrt{2\lambda_n}}\right)\right]\left(1 + \frac{y'^2}{2\lambda_n}\right)\exp\left(-\frac{y'^2}{2}\right)H_{2n+1}(y'),$$
$$G_n(\lambda_n) = \frac{c_n}{\sqrt{\lambda_n}}\int_0^\infty dy'\left(1 + \frac{y'^2}{2\lambda_n}\right)\exp\left(-\frac{y'^2}{2}\right)H_{2n+1}(y').$$

(19)

The quantities $R_0$ and $R_1$ encapsulate all the information concerning the long wave-length electrostatic ITG modes that we require.

## 5. The Ion FLR Region: $\bar{k} \sim 1$

The purpose of this section is to determine the electromagnetic perturbation driven by the long wave-length ITG mode in the region of k-space where finite ion Larmor radius effects are important, which can in turn be matched to the short wave-length response in the electron layer.

But first, as an aside, we note that in conventional tearing mode theory [12] one matches the perturbed electrostatic potential $\hat{\varphi}$ to an external ideal MHD form at low $\bar{k}$:

$$\hat{\varphi} \sim 1 + \frac{\Delta' \rho_i}{\pi \bar{k}}, \tag{20}$$

where we recall that in ideal MHD $\rho_i \Delta' = -2\bar{k}_y \equiv -2\sqrt{2b_0}$ for high wave numbers of stable tearing modes. Now, however, because of the ion dynamics it is the solution (15) that provides the boundary condition at low $\bar{k}$ for determining the non-local $\varphi^{(1)}$, which we now calculate. Thus, we must match to the form

$$\hat{\varphi} \sim \frac{1}{\bar{k}^2}\left(1 + \frac{\Delta^*}{\bar{k}}\right), \tag{21}$$

where

$$\Delta^* = \frac{R_1}{R_0} = -\frac{2\sqrt{2b_0}}{\pi} \frac{\left\{c_0 - \frac{i\bar{\omega}_n \hat{\beta}_T b_0^{5/2}}{\alpha\tau} G_n(\lambda_n)\varphi_0\right\}}{\left\{c_0 - \frac{i\bar{\omega}_n \hat{\beta}_T b_0^{5/2}}{\alpha\tau} F_n(\lambda_n)\varphi_0\right\}}. \tag{22}$$

We return to the calculation of $\hat{\varphi}^{(1)}$ in the ion FLR region where $\bar{k} \sim 1 \gg \bar{k}_y \sim \sqrt{2b_0}$ and the ion sound effects (i.e. terms involving $b_0 d^2/d\bar{k}^2$) are negligible. Equations (1) and (4) then simplify considerably to yield an equation for $\hat{\varphi}^{(1)}$:

$$\frac{d}{d\bar{k}}\bar{k}_\perp^2 \frac{d}{d\bar{k}}\left[(\tau H_0(\bar{k}_\perp) - 1)\hat{\varphi}^{(1)}\right] = 0, \tag{23}$$

where

$$H_0 = \Gamma_0 - 1 + \frac{\alpha \bar{k}_\perp^2}{2b_0 \bar{\omega}_n}(\Gamma_0 - \Gamma_1). \tag{24}$$

This has the solution

$$(\tau H_0(\bar{k}_\perp) - 1)\hat{\varphi}^{(1)} = \frac{c_0'}{\bar{k}_y}\tan^{-1}\left(\frac{\bar{k}}{\bar{k}_y}\right) + c_1', \tag{25}$$

so that, for $\bar{k}_y \ll \bar{k} \ll 1$,

$$\hat{\varphi}^{(1)} \sim \frac{2b_0\bar{\omega}_n}{\alpha\tau}\left[\frac{c'_0}{\bar{k}_y}\left(\frac{\pi}{2}-\frac{\bar{k}_y}{\bar{k}}\right)+c'_1\right]\frac{1}{\bar{k}^2}. \tag{26}$$

Matching to solution (15) and recalling the form (18) for $R(\bar{k})$ when $\bar{k}\ll 1$, we obtain equations that determine $c'_0$ and $c'_1$:

$$\frac{\pi}{2}\frac{c'_0}{\bar{k}_y}+c'_1=R_0, \quad c'_0=-R_1. \tag{27}$$

In the region $\bar{k}\gg 1$, where $H_0(\bar{k})\sim -1+\Lambda/\bar{k}$ with $\Lambda=(1+\alpha/2b_0\bar{\omega}_n)/\sqrt{\pi}$, the solution (25) takes the asymptotic form on making the substitutions (27):

$$\hat{\varphi}^{(1)}=-\frac{1}{(1+\tau)}\left(R_0+\left(R_1+\frac{\Lambda R_0\tau}{(1+\tau)}\right)\frac{1}{\bar{k}}\right), \tag{28}$$

To next order in $\hat{\beta}_T b_0^2$, eqn. (23) becomes

$$\frac{d}{d\bar{k}}\bar{k}_\perp^2\frac{d}{d\bar{k}}\left[(\tau H_0(\bar{k}_\perp)-1)\hat{\varphi}^{(2)}\right]+\frac{\bar{\omega}_n^2\hat{\beta}_T b_0^2}{\tau\alpha^2}H_0(\bar{k}_\perp)\hat{\varphi}^{(1)}=0 \tag{29}$$

and we find the correction $\varphi^{(2)}$, which in the limit $\bar{k}\gg 1$ takes the form:

$$\hat{\varphi}^{(2)}=\frac{\bar{\omega}_n^2\hat{\beta}_T b_0^2}{\alpha^2\tau(1+\tau)^2}\left(R_0\ell n\bar{k}+\frac{(\Lambda R_0-R_1)}{\bar{k}}\ell n\bar{k}+0\left(\frac{1}{\bar{k}}\right)\right). \tag{30}$$

Introducing the local ITG contribution $\hat{\varphi}_{ITG}$ and the non-local contributions $\hat{\varphi}^{(1)}$ and $\hat{\varphi}^{(2)}$ into eqn. (1) provides an equation for $\hat{A}(\bar{k})$ (retaining terms up to two orders in $\hat{\beta}_T b_0^2$) in the region $\bar{k}\gg 1$:

$$\bar{k}_\perp^2\hat{A}=c_0-i\frac{\bar{\omega}_n^2\hat{\beta}_T b_0^2}{\tau\alpha^2}\int_0^{\bar{k}}d\bar{k}\,H_0(\bar{k})\left(\hat{\varphi}_{ITG}+\hat{\varphi}^{(1)}+\hat{\varphi}^{(2)}\right). \tag{31}$$

At large $\bar{k}$ the integration over the ITG contribution, $\hat{\varphi}_{ITG}$, yields a constant, while for the non-local contributions, $\hat{\varphi}^{(1)}$ and $\hat{\varphi}^{(2)}$, one inserts the results (28) and (30) and uses the asymptotic form of $H_0(\bar{k})$ to obtain

$$\hat{A}(\bar{k})\sim -i\frac{R_1}{\bar{k}^2}\left(1+\frac{\bar{\omega}_n^2\hat{\beta}_T b_0^2}{\alpha^2\tau(1+\tau)}\ell n\bar{k}\right)-i\frac{\bar{\omega}_n^2\hat{\beta}_T b_0^2}{\alpha^2\tau(1+\tau)}\frac{R_0}{\bar{k}}\left(1-\frac{\bar{\omega}_n^2\hat{\beta}_T b_0^2}{\alpha^2\tau(1+\tau)}\ell n\bar{k}-\frac{\Lambda}{(1+\tau)}\frac{\ell n\bar{k}}{\bar{k}}\right), \tag{32}$$

where we have retained logarithmic corrections in order to match fully to the electron region solution. [The integral over $\bar{k}$ of $\hat{\varphi}^{(1)}$ also produces a term proportional to $\bar{k}^{-2}$, but this is smaller by order $\hat{\beta}_T b_0^2$ than the term retained in eqn. (32).]

We shall see that the same term involving $\Lambda$ is present in the electron region solution and so matches automatically. It is then helpful to recognise that eqn. (32) is the low $\hat{\beta}_T b_0^2$ expansion of an expression involving two power laws:

$$\hat{A}(\bar{k}) \sim -iR_1 \bar{k}^{-(\nu_+ + 1)} - i \frac{\bar{\omega}_n^2 \hat{\beta}_T b_0^2}{\alpha^2 \tau (1+\tau)} R_0 \bar{k}^{-(\nu_- + 1)}, \qquad (33)$$

where

$$\nu_\pm = \frac{1}{2} \pm \left( \frac{1}{4} - \frac{\bar{\omega}_n^2 \hat{\beta}_T b_0^2}{\tau(1+\tau)\alpha^2} \right)^{1/2}. \qquad (34)$$

We see that the quantity $\Delta^* = R_1 / R_0$ in eqn. (33) continues to play a key role in the matching condition.

## 6. The Electron Layer: $\bar{k} \tilde{\ } \rho_i / \delta_0$

To investigate the extent of anomalous transport due to the magnetic field perturbation associated with the electrostatic ITG mode we must relate the magnitude of the magnetic perturbation at the resonant surface, $x = 0$, to the amplitude of the ITG mode. This involves solving for the perturbed magnetic potential in the electron region of k–space, $\bar{k} \sim \rho_i / \delta_0$, and then matching the resulting expression in the limit $\bar{k} \delta_0 / \rho_i \ll 1$ to the form (33) from the ion region. In this section we obtain the electron region solution satisfying the correct boundary condition at $x = 0$.

In the limit $\bar{k} \gg 1$ we can solve eqn (1) for $\hat{\varphi}(\bar{k})$ in terms of $\hat{A}(\bar{k})$:

$$\hat{\varphi} = -\frac{i\tau\alpha^2}{\bar{\omega}^2 \hat{\beta}_T b_0^2} \left(1 - \frac{\Lambda}{\bar{k}}\right) \frac{d}{d\bar{k}} \left(\bar{k}_\perp^2 \hat{A}\right), \qquad (35)$$

where we have introduced the large $\bar{k}$ expansion of $H_0(\bar{k})$. After simplifying eqn. (2) in the limit $x \sim \delta_0 \ll \rho_i$, and substituting for $\hat{\varphi}(\bar{k})$ from eqn. (35), ignoring the correction from the term $\Lambda / \bar{k}$, we obtain an equation for $\hat{A}$ in x-space:

$$\frac{\tau^2\alpha^2}{\overline{\omega}_n^2\hat{\beta}_T b_0^2}\frac{d^2}{ds^2}\hat{A} = -\frac{(\hat{\sigma}_0 + d_1 s^2)}{(1 + d_0 s^2 + d_1 s^4)}\left(\hat{A} + \frac{\tau\alpha^2 s^2}{\overline{\omega}_n^2\hat{\beta}_T b_0^2}\frac{d^2}{ds^2}\hat{A}\right), \qquad (36)$$

where we have introduced $s = x/\delta$, i.e. we have scaled x to the complex electron semi-collisional width. Equation (36) can be re-arranged to yield

$$\frac{d^2}{ds^2}\hat{A} = -\frac{\overline{\omega}_n^2\hat{\beta}_T b_0^2}{\tau^2\alpha^2}\hat{\sigma}(s^2)\hat{A}, \qquad (37)$$

where

$$\hat{\sigma}(s^2) = \frac{\hat{\sigma}_0 + d_1 s^2}{1 + (d_0 + \hat{\sigma}_0/\tau)s^2 + d_1(1+\tau)s^4/\tau}. \qquad (38)$$

Invoking the low $\hat{\beta}_T b_0^2$ expansion we can iterate the solution of eqn. (36):

$$\hat{A} = A_0\left[1 - \frac{\overline{\omega}_n^2\hat{\beta}_T b_0^2}{\tau^2\alpha^2}\int_0^s ds'\int_0^{s'} ds''\hat{\sigma}(s''^2)\right]. \qquad (39)$$

Recalling that our eventual aim is to calculate A(0), the value of A at the resonant surface, s = 0, we need to determine the coefficient $A_0$ in eqn. (39).

The repeated integral in eqn. (39) can be evaluated straightforwardly and the limit $s \to \infty$ taken. On exploiting the condition $\hat{\sigma}_0 \gg 1$, this leads to

$$\hat{A} = A_0\left\{1 - \frac{\overline{\omega}_n^2\hat{\beta}_T b_0^2}{\tau^2\alpha^2}\left[\frac{\pi\sqrt{\tau\hat{\sigma}_0}}{2}s - \frac{\tau}{1+\tau}\ell n\,s\right]\right\}, \qquad s \to \infty \qquad (40)$$

We must relate this solution to its transform in k-space to link the amplitude $A_0$ to the quantity $c_0$ introduced in eqn. (11). In order to do this, it is convenient to consider the asymptotic form of the solution of eqn. (37) for finite values of $\hat{\beta}_T b_0^2$:

$$\hat{A} = b_+ s^{\nu_+} + b_- s^{\nu_-}; \quad \nu_\pm = \frac{1}{2} \pm \left(\frac{1}{4} - \frac{\overline{\omega}_n^2\hat{\beta}_T b_0^2}{\tau(1+\tau)\alpha^2}\right)^{1/2}. \qquad (41)$$

Now taking the limit $\hat{\beta}_T b_0^2 \ll 1$ we have

$$\hat{A} \to b_+ s + b_-\left(1 + \frac{\overline{\omega}_n^2\hat{\beta}_T b_0^2}{\tau(1+\tau)\alpha^2}\ell n\,s\right), \qquad (42)$$

so that we can identify $b_+$ and $b_-$ by comparing expressions (40) and (42). Thus

$$b_+ = -\frac{\pi}{2}\frac{\sqrt{1.71}\hat{\beta}_T}{\tau}\left(\frac{L_{Ti}}{L_{Te}}\right)^{1/2}\left(\frac{\overline{\omega}_n b_0}{\alpha}\right)^{3/2} A_0 \qquad (43)$$

[where we have substituted for $\hat{\sigma}_0$ from its definition in eqn. (3)], so that $A_0$ is determined by the value of $b_+$. Furthermore,

$$\frac{b_+}{b_-} = -\frac{\pi}{2}\frac{\sqrt{1.71}\hat{\beta}_T}{\tau}\left(\frac{L_{Ti}}{L_{Te}}\right)^{1/2}\left(\frac{\overline{\omega}_n b_0}{\alpha}\right)^{3/2}. \qquad (44)$$

The form (41) can be used to calculate the transform of $\hat{A}$ in $\overline{k}$-space [recalling $\hat{A}(s)$ is an even function of s [9]):

$$\hat{A}(t) = b_+ t^{-(\nu_+ + 1)}\cos\left[(\nu_+ + 1)\frac{\pi}{2}\right]\Gamma(\nu_+ + 1) + b_- t^{-(\nu_- + 1)}\cos\left[(\nu_- + 1)\frac{\pi}{2}\right]\Gamma(\nu_- + 1), \qquad (45)$$

where $t = (\delta/\rho_i)\overline{k}$. Taking the limit $\hat{\beta}_T b_0^2 \ll 1$, we obtain

$$\hat{A}(t) \simeq -b_+ t^{-(\nu_+ + 1)} - \frac{\pi}{2}\frac{\overline{\omega}_n^2 \hat{\beta}_T b_0^2}{\alpha^2 \tau(1+\tau)} b_- t^{-(\nu_- + 1)}. \qquad (46)$$

We can iterate on this solution to calculate the correction arising from the $\Lambda/\overline{k}$ term in eqn. (35) and find it is identical to the related term in the ion region eqn. (32). The form of the solution (46) thus has an identical structure to that in eqn. (33) and the two can therefore be matched asymptotically.

## 7. Matching Electron and Ion Region Solutions

In this section we first obtain the constant $c_0$ in terms of the ITG mode amplitude $\varphi_0$ by matching the electron and ion region solutions in the region $\overline{k} \gg 1$. We can then fully determine the amplitude of the magnetic perturbation at the resonant surface, $x = 0$.

To achieve this, the expression (46) must be matched to the large k limit of the ion region solution, namely eqn. (33), to obtain:

$$\frac{\delta}{\rho_i}\frac{R_1}{R_0} = \frac{2}{\pi}\frac{b_+}{b_-} \qquad (47)$$

After substituting eqns. (22) and (44) we find

$$c_0 = i\frac{\overline{\omega}_n b_0^{1/2}}{\alpha\tau}\frac{\dfrac{\delta}{\rho_i}G_n(\lambda_n) - \dfrac{\pi}{2\tau}\left(\dfrac{1.71}{2}\right)^{1/2}\left(\dfrac{\overline{\omega}_n}{\alpha}\right)^{3/2}\left(\dfrac{L_{Ti}}{L_{Te}}\right)^{1/2}b_0\hat{\beta}_T F_n(\lambda_n)}{\dfrac{\delta}{\rho_i} - \dfrac{\pi}{2\tau}\left(\dfrac{1.71}{2}\right)^{1/2}\left(\dfrac{\overline{\omega}_n}{\alpha}\right)^{3/2}\left(\dfrac{L_{Ti}}{L_{Te}}\right)^{1/2}b_0\hat{\beta}_T}\hat{\beta}_T b_0^2 \varphi_0, \quad (48)$$

where we have ignored terms of order $\overline{\omega}_n^2 \hat{\beta}_T^2 b_0^2 (\ell n(\delta/\rho_i) + \gamma - 1)/\alpha^2 (1+\tau)\tau \ll 1$, $\gamma = 0.57721\ldots$ being the Euler-Mascheroni constant. As noted above, the term arising from the $\Lambda/\overline{k}$ corrections to H continues smoothly from the ion to electron regions and played no part in the matching condition.

In the stability theory of semi-collisional tearing modes [9], such a matching condition determines the eigenfrequency $\omega$, whereas in the present calculation it determines the amplitude $c_0$, the frequency having already been determined by the earlier ITG mode analysis.

In order to determine the magnetic perturbation at x = 0, we require $A(0) = A_0$. This can be obtained in terms of $b_+$ through eqn. (43) with $b_+$ in turn determined by matching corresponding powers in eqns. (33) and (46). Thus

$$b_+ = -\left(\frac{\delta}{\rho_i}\right)^2 \left(c_0 - i\frac{\overline{\omega}_0 \hat{\beta}_T b_0^{5/2}}{\alpha\tau}G_n(\lambda_n)\varphi_0\right) \qquad (49)$$

so that

$$A_0 = -\frac{2}{\pi}\frac{\tau}{\sqrt{1.71}\hat{\beta}_T}\left(\frac{L_{Te}}{L_{Ti}}\right)^{1/2}\left(\frac{\alpha}{\overline{\omega}_n b_0}\right)^{3/2}\left(\frac{\delta}{\rho_i}\right)^2\left(c_0 - i\frac{\overline{\omega}_n b_0^{1/2}}{\alpha\tau}G_n(\lambda_n)\hat{\beta}_T b_0^2 \varphi_0\right) \qquad (50)$$

and finally

$$A(0) = \frac{i}{\sqrt{2}}\frac{\overline{\omega}_n}{\alpha\tau}\left(\frac{\delta}{\rho_i}\right)^2 \frac{K_n(\lambda_n)}{\dfrac{\delta}{\rho_i} - \dfrac{\pi}{2}\sqrt{\dfrac{1.71}{2}}\dfrac{b_0\hat{\beta}_T}{\tau}\left(\dfrac{L_{Ti}}{L_{Te}}\right)^{1/2}\left(\dfrac{\overline{\omega}_n}{\alpha}\right)^{3/2}} b_0^2 \hat{\beta}_T \varphi_0 \qquad (51)$$

where

$$K_n(\lambda_n) = G_n(\lambda_n) - F_n(\lambda_n) = \frac{c_n}{\sqrt{\lambda_n}}\frac{2}{\pi}\int_0^\infty dy' \tan^{-1}\left(\frac{y'}{\sqrt{2\lambda_n}}\right)\left(1 + \frac{y'^2}{2\lambda_n}\right)\exp\left(-\frac{y'^2}{2}\right)H_{2n+1}(y')$$
$$(52)$$

encapsulates the effects of the long wave-length electrostatic ITG mode.

## 8. The Tearing Parity Magnetic Perturbation

There are two interesting limits to consider: (i) $|\bar{\omega}_n| b_0 \hat{\beta}_T / \alpha^{3/2} \ll \delta_0/\rho_i$ and and (ii) $\delta_0/\rho_i \ll |\bar{\omega}_n| b_0 \hat{\beta}_T / \alpha^{3/2}$ where we recall $\delta_0 = |\delta(\bar{\omega}=1)|$ with $\delta \propto \bar{\omega}^{1/2}$. In the lower $\hat{\beta}_T$ case (i), we have

$$A_0 = \frac{\exp(i\pi/4)}{\sqrt{2}\tau} \hat{\beta}_T b_0^2 \left(\frac{\delta_0}{\rho_i}\right)\left(\frac{\bar{\omega}_n^{3/2}}{\alpha}\right) K_n(\lambda_n)\varphi_0, \qquad (53)$$

whereas in the higher $\hat{\beta}_T$ case (ii):

$$A_0 = -\frac{2b_0}{\pi\sqrt{1.71}}\left(\frac{L_{Te}}{L_{Ti}}\right)^{1/2}\left(\frac{\delta_0}{\rho_i}\right)^2 \alpha^{1/2} \bar{\omega}_n^{1/2} K_n(\lambda_n)\varphi_0. \qquad (54)$$

Now

$$\frac{\delta B_x(0)}{B} \sim k_y \frac{A_\parallel(0)}{B} \sim k_y \frac{b_0^{1/2} T_i}{\omega e B L_s} A(0) = \frac{b_0^{3/2}}{\alpha^{1/2}\bar{\omega}_n} A(0), \qquad (55)$$

so that at lower $\hat{\beta}_T$

$$\frac{\delta B_x(0)}{B} \sim \left(\frac{\delta_0}{\rho_i}\right)\frac{\hat{\beta}_T b_0^{7/2} \bar{\omega}_n^{1/2}}{\tau\alpha^{3/2}} K_n(\lambda_n)\frac{e\varphi_0}{T_i}, \qquad (56)$$

while at higher $\hat{\beta}_T$

$$\frac{\delta B_x(0)}{B} \sim \left(\frac{\delta_0}{\rho_i}\right)^2 \frac{b_0^{5/2}}{\bar{\omega}_n^{1/2}}\left(\frac{L_{Te}}{L_{Ti}}\right)^{1/2} K_n(\lambda_n)\frac{e\varphi_0}{T_i}, \qquad (57)$$

where we recall $\bar{\omega}_n$ and $\lambda_n$ also depend on $\alpha = (b_0^2 L_s/L_T)^{2/3}$ and $n$.

We can use expressions (56) and (57) to explore the relative contributions of the higher n modes, for which $\lambda_n \gg 1$. In Fig. 1 we plot the real and imaginary parts of $K_n$ as a function of n for the value of $\lambda$ corresponding to $\alpha = \tau = 1$, finding that $K_n(\lambda_n) \sim$ constant for $n \gg 1$. Recalling $\bar{\omega}_n \sim (2n+1)^{1/2}$ and $\lambda_n \sim (2n+1)^{1/2}$, we deduce that, at lower $\hat{\beta}_T$, $\delta B_x(0)/B \propto n^{1/4}\hat{\beta}_T$, so that the linear slope with $\hat{\beta}_T$ becomes steeper. At higher $\hat{\beta}_T$, $\delta B_x(0)/B \propto n^{-1/4}$, so that the perturbed magnetic field is smaller. The transition between the lower and higher $\hat{\beta}_T$ cases scales as $\hat{\beta}_T \propto n^{-1/2}$ so that the higher $\hat{\beta}_T$ result has a lower threshold in $\hat{\beta}_T$, reconciling these two scalings. However, at the lower values of $\hat{\beta}_T$ there is a finite value of n, which increases as $\hat{\beta}_T$ decreases, where $\delta B_x(0)/B$ is a maximum.

These features are evident in the results of Fig. 2, where the variation of $\delta B_x(0)/B$ (normalised to $e\varphi_0/T_i$) with $\hat{\beta}_T$, calculated from eqn. (55) but using the full expression (51) for $A(0)$, is shown for a number of n values. We take representative values of the other parameters to be $\delta_0/\rho_i = 0.1$, $b_0 = 0.1$, $\alpha = \tau = 1$ and $L_{Ti} = L_{Te}$. Finally Fig. 3 presents a three-dimensional visualisation of the surface of $\delta B_x(0)/B$ in the space of $\hat{\beta}_T$ and n for the same set of the parameters. It can be seen that the variation in amplitude of $\delta B_x(0)/B$ with n is not great and we can therefore estimate the amount of magnetic reconnection by considering just the low n modes.

We will, therefore, specialise to the case n = 0, corresponding to the lowest odd-k mode. In this case we can evaluate $K_0(\lambda_0)$ as [13]

$$K_0(\lambda_0) = \frac{1}{\pi^{1/4}\sqrt{2\lambda_0}}\left[\frac{1}{\sqrt{\pi\lambda}} + \frac{e^\lambda}{\lambda}\left(1 - \text{erf}\left(\sqrt{\lambda}\right)\right)\right] \tag{58}$$

To express our results in more physically meaning parameters we recall that $\delta_0/\rho_i = b_0^{-5/12}(m_e/2m_i)^{1/4}(T_i/T_e)^{1/2}(\nu_{ei}L_s/v_{the})^{1/2}(L_s/L_{Ti})^{1/6}$; it is thus useful to extract the dependencies on the wave-number $b_0$ and $\tau$, writing $\delta_0/\rho_i = b_0^{-5/12}\tau^{-1/4}\hat{\delta}$, i.e. $\hat{\delta} = 2^{-3/2}(m_e/m_i)^{1/4}(\nu_{ei}L_s/v_{the})^{1/2}(L_s/L_{Ti})^{1/6}$.

We can consider two additional limits: (i) $\alpha \ll 1$, and (ii) $\alpha \gg 1$, using the results in eqns. (9) and (10) for n = 0. In case (i) $\bar{\omega}_0 \sim \alpha^{1/4}\tau^{1/2}$ and $\lambda_0 \sim \alpha^{3/4}\tau^{1/2}$; then eqn. (58) implies $K \sim \lambda_0^{-3/2}\left(1 - \sqrt{\lambda_0/\pi}\right) \sim \alpha^{-9/8}\tau^{-3/4}\left(1 - \alpha^{3/8}\tau^{1/4}/\sqrt{\pi}\right)$ since $\lambda_0 \ll 1$. Thus at lower $\hat{\beta}_T$ we obtain

$$\frac{\delta B_x(0)}{B} \sim \frac{b_0^{-1/4}\hat{\beta}_T\hat{\delta}}{\tau^{7/4}}\left(\frac{L_{Ti}}{L_s}\right)^{5/3}\frac{e\varphi_0}{T_i}, \tag{59}$$

while for higher $\hat{\beta}_T$ we have

$$\frac{\delta B_x(0)}{B} \sim \frac{\hat{\delta}^2}{\tau^{3/2}} \left(\frac{L_{Te}}{L_{Ti}}\right)^{1/2} \left(\frac{L_{Ti}}{L_s}\right)^{5/6} \frac{e\varphi_0}{T_i} \quad . \tag{60}$$

In case (ii) $\bar{\omega}_0 \sim \alpha\tau$ and $K_0 \sim \lambda_0^{-1} \sim \alpha^{-3/2}\tau^{-1}$ so that at lower $\hat{\beta}_T$ we find

$$\frac{\delta B_x(0)}{B} \sim \frac{b_0^{-1/4} \hat{\beta}_T \hat{\delta}}{\tau^{7/4}} \left(\frac{L_{Ti}}{L_s}\right)^{5/3} \frac{e\varphi_0}{T_i}, \tag{61}$$

identical to result (59), and at higher $\hat{\beta}_T$

$$\frac{\delta B_x(0)}{B} \sim \frac{\hat{\delta}^2}{b_0 \tau^2} \left(\frac{L_{Te}}{L_{Ti}}\right)^{1/2} \left(\frac{L_{Ti}}{L_s}\right)^{4/3} \frac{e\varphi_0}{T_i} \quad . \tag{62}$$

The lower $\hat{\beta}_T$ results for the magnetic perturbations, eqns. (59) and (61), decrease weakly with increasing $b_0$, whereas the higher $\hat{\beta}_T$ ones approach a constant (with a $b_0^{1/2}$ correction) for $\alpha < 1$, i.e. at longer wave-length, and fall rapidly when $\alpha > 1$. The higher $\hat{\beta}_T$ result (60) and result (62) match at $b_0 = (L_{Ti}/\tau L_s)^{1/2}$, i.e. at $\alpha = 1$ for $\tau = 1$. This value is given by eqn. (60) and corresponds to the situation with the largest magnetic perturbation when regarded as a function of $\hat{\beta}_T$ and $b_0$. A numerical calculation of the variation of $\delta B_x(0)/B$ with $\hat{\beta}_T$ and $b_0$ for $n = 0$, $\hat{\delta} = 0.005$, $L_s/L_{Ti} = 25$, $L_{Ti} = L_{Te}$ and $\tau = 1$, again calculated from eqn. (55) and using the full expression (51) for A(0), is shown in Fig. 4: it illustrates the various features described above.

## 9. Stochastic Field Transport

In this section we use the values obtained for the magnetic perturbation at the resonant surface to estimate the resulting stochastic magnetic field transport and compare it with the basic $E \times B$ transport due to the original electrostatic ITG mode.

These are achieved by first inserting expression (60) for the largest 'radial' magnetic field perturbation, $\delta B_x / B$, into the Rechester - Rosenbluth formula [10] for the stochastic magnetic field contribution to the electron thermal diffusivity:

$$\chi_e \sim v_{the} \left(\frac{\delta B_x}{B}\right)^2 L, \tag{63}$$

where $v_{the}$ is the electron thermal velocity and $L$ is the correlation length along the magnetic field of the magnetic perturbation. Taking $L \sim k_\parallel^{-1} \sim \left(k_y \delta_0 / L_s\right)^{-1}$, we estimate the magnitude of expression (63). The result is

$$\chi_e \sim \hat{\delta}^3 \frac{v_{the} L_s}{b_0^{1/12} \tau^{11/4}} \frac{L_{Te}}{L_{Ti}} \left(\frac{L_{Ti}}{L_s}\right)^{5/3} \left(\frac{e\varphi_0}{T_i}\right)^2. \tag{64}$$

It can be seen that it is a very weak function of $b_0$ but of course depends on $\varphi_0$, the amplitude of the ITG mode. For orientation, we can make a tentative estimate for this using a mixing-length model for the non-linear saturation of the ITG mode:

$$\frac{e\varphi}{T_i} = c \frac{1}{k_y L_{Ti}} \quad, \tag{65}$$

with c some constant, leading to the result

$$\chi_e^{st} \sim c^2 v_{thi} R \frac{\rho_i^2}{L_{Ti}^2} \frac{\varepsilon^{9/4} v_{*e}^{3/2}}{q^{1/6} s^{4/3} b_0^{13/12} \tau^{11/4}} \left(\frac{m_e}{m_i}\right)^{1/4} \left(\frac{L_{Te}}{L_{Ti}}\right) \left(\frac{L_{Ti}}{R}\right)^{7/6}, \tag{66}$$

where we have introduced the standard tokamak parameters: collisionality, $v_{*e} = v_{ei} Rq / \varepsilon^{3/2} v_{the}$ with $\varepsilon = r/R$, the inverse aspect ratio of a torus, q the safety factor and the magnetic shear, $s = (r/q) dq/dr$, since $L_s = Rq/s$. Expression (66) has a gyro-Bohm scaling, but with a significant dependence on $b_0$ and an unknown amplitude, c.

It is interesting to compare the island width, $w = \left(8(L_s/k_y)\delta B_x/B\right)^{1/2}$ associated with the magnetic perturbation (60) with the reconnection width, $\delta_0$. Using the mixing-length like estimate (63) and eqn. (60) for $\delta B_x$, we find that

$$\frac{w}{\delta_0} \sim \frac{c^{1/2}}{\tau^{1/2}} \left(\frac{L_{Te}}{L_{Ti}}\right)^{1/4} \left(\frac{L_s}{b_0 L_{Ti}}\right)^{1/12}. \tag{67}$$

This implies that the linear theory is only valid if the factor c in the saturation level (65) for the ITG electrostatic potential satisfies the condition

$$c < \tau \frac{L_{Ti}}{L_{Te}} \left(\frac{b_0 L_{Ti}}{L_s}\right)^{1/6} \quad, \tag{68}$$

limiting the magnitude of the estimate (66).

However, since the more fundamental expression (64) depends on the magnitude of $\varphi_0$, it is perhaps more meaningful to compare it with an estimate of the actual $\delta E \times B$ (where $\delta E_y = -ik_y \varphi$) cross-field transport due to the electrostatic ITG perturbation. Assuming a correlation time $\omega_0^{-1}$ with step-lengths $\delta E_y / \omega_0 B$

$$\chi_e^{E \times B} \sim \left(\frac{\delta E_y}{B}\right)^2 \frac{1}{\omega_0} \sim b_0 \frac{v_{thi}^2}{\omega_0} \left(\frac{e\varphi_0}{T_i}\right)^2. \tag{69}$$

Thus the ratio of $E \times B$ transport to stochastic magnetic field transport becomes

$$\frac{\chi_e^{st}}{\chi_e^{E \times B}} \sim \frac{\hat{\delta}^3}{(b_0^{1/3}\tau)^{11/4}} \left(\frac{m_i}{m_e}\right)^{1/2} \left(\frac{L_{Te}}{L_{Ti}}\right) \left(\frac{L_{Ti}}{L_s}\right)^{4/3}, \tag{70}$$

where we have taken $\bar{\omega}_0 = 1$ and $\tau = 1$. This can be expressed in the form

$$\frac{\chi_e^{st}}{\chi_e^{E \times B}} \sim \frac{\varepsilon^{9/4} v_{*e}^{3/2}}{s^{2/3} q^{5/6} (b_0^{1/3}\tau)^{11/4}} \left(\frac{m_e}{m_i}\right)^{1/4} \left(\frac{L_{Te}}{L_{Ti}}\right) \left(\frac{L_{Ti}}{R}\right)^{5/6}. \tag{71}$$

This clearly increases as $b_0$ decreases, but we require $b_0 > \hat{\delta}^{12/5}$ to ensure that the validity condition $\delta_0 / \rho_i < 1$ is satisfied. This provides an upper bound

$$\frac{\chi_e^{st}}{\chi_e^{E \times B}} < \frac{\varepsilon^{3/5} s^{4/5} v_{*e}^{2/5}}{q^{6/5} \tau^{11/4}} \left(\frac{m_i}{m_e}\right)^{3/10} \left(\frac{L_{Te}}{L_{Ti}}\right) \left(\frac{L_{Ti}}{R}\right)^{6/5}. \tag{72}$$

We note that consideration of the higher n modes leads to the result $\chi_e^{st} / \chi_e^{E \times B} \sim n^{-1/4}$, confirming the importance of lower n values.

## 10. Conclusions

To explore the possibility that the electromagnetic corrections to an essentially electrostatic instability occurring at small, but finite β can produce significant stochastic magnetic field transport, we have examined the case of the odd-parity, long wave-length ITG modes in sheared plasma slab geometry, with semi-collisional electron physics to permit reconnection.

The calculation resembles that of the finite ion Larmor radius effect on tearing mode instability [9] where the $\Delta'$ instability drive acts as a boundary condition at long radial wave-lengths. In this case, using such a boundary condition one calculates the ion response in the shorter wave-length region where finite ion Larmor radius effects are important. Finally, the solution in this region is matched to that in the short wave-length region where electron dissipative effects dominate and permit magnetic reconnection to occur. This matching condition provides a dispersion relation for the tearing mode.

In the present case the electrostatic ITG dynamics is found to provide a modified long wave-length boundary condition valid in the region $k_y\rho_i \ll k\rho_i \ll 1$. This involves a quantity $\Delta^*$, characteristic of the ITG mode. The boundary condition is applied to the electromagnetic response in the ion region, $k\rho_i \sim 1$. This solution is then matched to the semi-collisional electron solution in the region $k\rho_i \sim \rho_i/\delta_0 \gg 1$. The matching provides an eigenvalue condition that determines the relative amplitude of the electromagnetic perturbation in the ion region driven by the electrostatic ITG mode. Although the higher n 'radial harmonics' of the ITG mode have greater growth rates, we find the lower values of n are more relevant for estimating the amount of reconnection.

However, since the ITG frequency differs from that of a tearing mode, the perturbed magnetic field it actually drives near the resonant surface, where reconnection takes place, is greatly diminished. Introducing a mixing-length like estimate for the saturated amplitude of the ITG mode leads to a gyro-Bohm scaling for $\chi_e^{st}$, though with a significant scaling with $b_0 = k_y^2\rho_i^2/2$: $\chi_e^{st} \sim b_0^{-12/11}$. It should be noted that our calculation has assumed that the electron layer is governed by linear physics, i.e. that the magnetic island associated with the magnetic perturbation is less than the semi-collisional width. For this to be valid the saturation level of the electrostatic ITG mode must be less than the mixing-length estimate, leading to a rather low value for $\chi_e^{st}$. A direct comparison of the Rechester-Rosenbluth estimate for the associated stochastic magnetic field transport with the $E \times B$ transport associated with the electrostatic ITG mode, indicates that their ratio, in standard tokamak parameters, is given by $\chi_e^{st}/\chi_e^{E\times B} \sim \varepsilon^{9/4} s^{-7/24} q^{-29/24} v_{*e}^{3/2} (m_e/m_i)^{1/4} (L_{Ti}/R)^{29/24}$, i.e., somewhat small. These findings are consistent with results from non-linear simulations of electromagnetic ITG turbulence [7] which found that non-linearly excited, linearly stable micro-tearing modes, rather than the unstable tearing parity ITG modes themselves, were responsible for the stochastic magnetic field transport observed in these calculations.

One can also envisage a similar calculation for other long wave-length modes such as the electrostatic trapped electron mode. This would produce an expression for $\Delta^*$ (the quantity encapsulating the influence of the electrostatic mode in driving reconnection, defined in eqn. (22)) that is characteristic of that mode rather than the ITG. Indeed, modes propagating in the same electron direction as a tearing mode may excite a greater electromagnetic response than the ITG mode. However, the separation of scales employed in the present calculation for the long wave-length ITG mode, namely $k\rho_i \ll 1$, allowing a distinction between the driving region of k-space and the full ion response when $k\rho_i \sim 1$, may become problematic and one would have to treat the whole ion region as one, before matching to the electron region.

# References

[1] J W Connor and H R Wilson, Plasma Phys. and Control Fusion **36** 719 (1994)
[2] J F Drake, N T Gladd, C S Liu and C L Chang, Phys. Rev. Lett. **44** 994 (1980)
[3] D J Applegate et al, Phys. Plasmas **11** 5085 (2004)
[4] W Guttenfelder et al, Phys. Plasmas **19** 056119 (2012)
[5] D Told et al, Phys. Plasmas **15** 102396 (2008)
[6] B Coppi, M N Rosenbluth and R Z Sagdeev, Phys. Fluids **10** 582 (1967)
[7] D R Hatch, M J Pueschel, F Jenko, W M Nevins, P W Terry and H Doerk, Phys. Plasmas **20** 012307 (2013)
[8] S C Cowley, R M Kulsrud and T S Hahm, Phys. Fluids **29** 3230 (1986)
[9] J W Connor, R J Hastie and A Zocco, Plasma Phys. Control. Fusion **54** 035003 (2012)
[10] A B Rechester and M N Rosenbluth, Phys. Rev. Lett. **40** 38 (1978)
[11] *Handbook of Mathematical Functions*, eds. M Abramowitz and I A Stegun, Dover Publications, Inc., New York, 1972, Chapter 19, Parabolic Cylinder Functions, J C P Miller
[12] F Pegoraro and T J Schep, Plasma Phys. Control. Fusion **28** 647 (1986)
[13] *Tables of Integrals, Series and Products*, I S Gradshteyn and I M Ryzhik, Academic Press Inc., San Diego, 1994

**Acknowledgements**

This work was funded by the RCUK Energy Programme [grant number EP/I501045] and the European Communities under the contract of Association between EURATOM and CCFE. To obtain further information on the data and models underlying this paper please contact PublicationsManager@ccfe.ac.uk. The views and opinions expressed herein do not necessarily reflect those of the European Commission. The authors gratefully acknowledge the contribution of Professor S C Cowley in initiating this work.

## Appendix A: Ion Region Equations

We require the ion response to the perturbed electromagnetic fields including full ion Larmor radius (FLR) effects and ion sound effects. As we consider the ITG mode in sheared slab geometry we adopt a co-ordinate system with x normal to the surfaces of constant density and temperature, z along the main magnetic field, $B_0$ and with y perpendicular to both of these. The y-component of the magnetic field varies linearly with x: $B_y = xB_0/L_s$, providing the magnetic shear with scale-length $L_s$. Perturbations have the form $\varphi \sim \varphi(x)\exp(ik_y y + ik_z z - i\omega t)$, where we measure x from the position where $k_\parallel(x) = 0$, i.e. $x_0 = -(k_z/k_y)L_s$. It is convenient to employ a Fourier representation to describe ion FLR effects so that $k_\parallel = k_y x/L_s \to -i(k_y/L_s)d/dk_x$ and the ion gyro-kinetic equation can be written

$$-i\omega h_i + \frac{k_y}{L_s} v_\parallel \frac{d}{dk_x} h_i = -i\frac{Ze}{T_j} F_{0i}(\omega - \omega_{*i}^T) J_0(z_i)(\varphi - v_\parallel A_\parallel) \;, \tag{A.1}$$

where the full ion distribution function has been expressed as

$$\delta f_i = -\frac{Ze\varphi}{T_i} F_{0i} + h_i e^{iL_i} \;, \tag{A.2}$$

with

$$L_i = |\mathbf{k} \times \mathbf{v}_\perp / \Omega_i|, \qquad z_i = \frac{k_\perp v_\perp}{\Omega_i}, \qquad \Omega_i = \frac{ZeB}{m_i},$$

$$\omega_{*i}^T = \omega_{*i}\left[1+\eta_i(u^2 - \frac{3}{2})\right], \; u^2 = \frac{m_i v^2}{2T_i}, \; \omega_{*i} = -\frac{k_y T_i}{Ze}\frac{d\ln n_i}{dr}, \eta_i = \frac{d(\ell n\, T_i)}{d(\ell n\, n_i)} \equiv \frac{L_n}{L_{Ti}}.$$

$$\tag{A.3}$$

Here $v_\parallel$ is the particle velocity along the magnetic field, $J_0$ is the zero order Bessel Functions, $F_{0i}$ is the Maxwellian distribution function, $\varphi$ is the perturbed electrostatic potential and $A_\parallel$ is the parallel component of the perturbed vector potential.

We solve eqn. (A.1) by utilising an ordering scheme appropriate to the slab ITG mode where in general we consider $\eta_i \sim 1$, introducing a small parameter, $\varepsilon = (L_{Ti}/L_s)^{1/3}$, such that $\omega/\omega_{*i} \sim \varepsilon^2$ and $k_\parallel v_{thi}/\omega \sim \varepsilon$. The solution is linear in $\varphi$ and $A_\parallel$ so the response to these two fields can be calculated independently. The lowest order response of $h_i$ to $\varphi$ contributes to the perturbed ion density, whereas the lowest order response to $A_\parallel$ produces a perturbed ion current. We therefore choose to define

$$h_i^{(0)} = -\frac{\omega_{*i}^T}{\omega} \frac{ZeF_{0i}}{T_i} J_0(z_i)\varphi, \tag{A.4}$$

while in first order we take

$$h_i^{(1)} = -iv_\parallel \frac{\omega_{*i}^T}{\omega^2} \frac{ZeF_{0i}}{T_i} \left[-\frac{k_y}{L_s}\frac{d}{dk_x}\left(J_0(z_i)\varphi\right) + i\omega J_0(z_i) A_\parallel\right], \tag{A.5}$$

so we see that our ordering scheme corresponds to comparable contributions of $\varphi$ and $A_\parallel$ to $E_\parallel$. Finally, in second order, we obtain

$$h_i^{(2)} = \frac{ZeF_{0i}}{T_i} J_0(z_i)\varphi - v_\parallel^2 \frac{\omega_{*i}^T}{\omega^3} \frac{ZeF_{0i}}{T_i}\left[-\frac{k_y^2}{L_s^2}\frac{d^2}{dk_x^2}\left(J_0(z_i)\varphi\right) + i\frac{k_y}{L_s}\omega\frac{d}{dk_x}\left(J_0(z_i)A_\parallel\right)\right] \tag{A.6}$$

yielding an electromagnetic contribution to the perturbed ion density.

Thus, the total perturbed ion density is given by the sum of the Boltzmann and non-adiabatic responses:

$$\frac{\delta n_i}{n_i} = -\frac{Ze\varphi}{T_i} + \frac{1}{n_i}\int d^3v J_0(z_i)\left(h_i^{(0)} + h_i^{(2)}\right), \tag{A.7}$$

for use in the quasi-neutrality condition which equates the perturbed ion and electron densities. In Appendix B we compute the perturbed electron density, which introduces the perturbed parallel electron velocity, $u_{\|e}$. It is more convenient to express this in terms of the perturbed parallel current $j_\|$, which we relate to $A_\|$ using Ampère's law, and the perturbed ion velocity, $u_{\|i}$, which is obtained from $h_i^{(1)}$:

$$u_{\|i} = \frac{1}{n_i}\int d^3v\, v_\| J_0(z_i)h_i^{(1)}. \tag{A.8}$$

Thus

$$u_{\|e} = u_{\|i} - \frac{j_\|}{n_e e} = u_{\|i} - \frac{k_\perp^2}{\mu_0 n_e e}A_\|. \tag{A.9}$$

Performing the integrations in eqns. (A.7) and (A.8), we find

$$\frac{\delta n_i}{n_i} = \frac{Ze\varphi}{T_i}\left[\left(1-\frac{\omega_{*i}}{\omega}\right)\Gamma_0 + \eta_i\frac{\omega_{*i}}{\omega}\frac{T_i k_\perp^2}{m_i \omega_{ci}^2}(\Gamma_0 - \Gamma_1) - 1\right]$$
$$+ \frac{\omega_{*i}}{\omega^3}\frac{k_y}{L_s}\frac{T_i}{m_i}\frac{Ze}{T_i}\left\{\frac{k_y}{L_s}\left[\alpha_1\frac{d^2\varphi}{dk_x^2} + \frac{d\varphi}{dk_x}\frac{d\alpha_1}{dk_x} - \frac{\varphi}{k_\perp^2}\left(\frac{2T_i k^2}{m_i \omega_{ci}^2}\alpha_2 + \frac{(2k_x^2 - k_\perp^2)}{2k_x}\frac{d\alpha_1}{dk_x}\right)\right]\right\}$$
$$+ \frac{\omega_{*i}}{\omega^3}\frac{k_y}{L_s}\frac{T_i}{m_i}\frac{Ze}{T_i}\left[-i\omega\left(\frac{A_\|}{2}\frac{d\alpha_1}{dk_x} + \alpha_1\frac{dA_\|}{dk_x}\right)\right]$$
-
$$\tag{A.10}$$

and

$$u_{\|i} = -i\frac{\omega_{*i}}{\omega^2}\frac{T_i}{m_i}\frac{Ze}{T_i}\left[-\frac{k_y}{L_s}\left(\frac{\varphi}{2}\frac{d\alpha_1}{dk_x} + \alpha_1\frac{d\varphi}{dk_x}\right) + i\omega\alpha_1 A_\|\right], \tag{A.11}$$

where

$$\alpha_1 = 2\int d^3v \frac{m_i v_\parallel^2}{2T_i}\left(1+\eta_i\left(\frac{m_i v^2}{2T_i}-\frac{3}{2}\right)\right)J_0^2\left(\frac{k_\perp v_\perp}{\omega_{ci}}\right)F_{Mi}\left(\frac{m_i v^2}{2T_i}\right)$$

(A.12)

$$\alpha_2 = 2\int d^3v \frac{m_i v_\parallel^2}{2T_i}\frac{m_i v_\perp^2}{2T_i}\left(1+\eta_i\left(\frac{m_i v^2}{2T_i}-\frac{3}{2}\right)\right)J_0^2\left(\frac{k_\perp v_\perp}{\omega_{ci}}\right)F_{Mi}\left(\frac{m_i v^2}{2T_i}\right),$$

so that

$$\alpha_1 = (1+\eta_i)\Gamma_0 + \eta_i b(\Gamma_1 - \Gamma_0) \equiv \bar{\alpha}_1 + \eta_i \hat{\alpha}_1$$

$$\alpha_2 = \frac{1}{2}\left[\Gamma_0 + b(\Gamma_1 - \Gamma_0)\right] + \eta_i\left[2(b-1)^2\Gamma_0 - b(2b-3)\Gamma_1\right] \equiv \bar{\alpha}_2 + \eta_i \hat{\alpha}_2,$$

(A.13)

with $\Gamma_n(b) = \exp(-b)I_n(b)$ and $b = k_\perp^2 \rho_i^2/2$.

## Appendix B: Electron Region Equations

In the semi-collisional regime we describe the electrons by the linearised Braginskii fluid equations: a continuity equation,

$$\omega \frac{\delta n_e}{n_e} = k_\parallel(x)u_{\parallel e} + \omega_{*e}\frac{e\varphi}{T_e},$$

(B.1)

and a parallel momentum equation

$$-i\omega\left[1-\frac{\omega_{*e}}{\omega}(1+1.71\eta_e)\right]A_\parallel + ik_\parallel(x)\varphi - 1.71ik_\parallel(x)\frac{\delta T_e}{e} - ik_\parallel(x)\frac{T_e}{e}\frac{\delta n_e}{n_e} = m_e\nu_{ei}\left(u_{\parallel e} - u_{\parallel i}\right),$$

(B.2)

where the perturbed electron temperature, $\delta T_e$, is given by the electron energy equation

$$\left(\frac{3}{2}\omega + i\kappa_{\parallel e}k_\parallel^2(x)\right)\frac{\delta T_e}{T_e} = 1.71k_\parallel(x)u_{\parallel e} + i\eta_e\omega_{*e}\kappa_{\parallel e}k_\parallel(x)A_\parallel + \frac{3}{2}\omega_{*e}\eta_e\frac{e\varphi}{T_e},$$

(B.3)

with $\kappa_{\parallel e} = 3.2T_e/m_e\nu_{ei}$ the parallel electron thermal diffusivity.

We solve for $u_{\parallel e}$ from eqns. (B.1) - (B.3):

$$u_{\|e} = -\frac{e}{m_e \nu_{ei}} \frac{\left[\sigma_0 + 2.1\sigma_1 \left(i k_\|^2 T_e / m_e \nu_{ei} \omega\right)\right]}{\left[1 + 5.1\left(i k_\|^2 T_e / m_e \nu_{ei} \omega\right) + 2.1\left(i k_\|^2 T_e / m_e \nu_{ei} \omega\right)^2\right]} \left[i\omega A_\| - i k_\|(x)\varphi\right]$$
$$+ \frac{\left[1 + 2.1\left(i k_\|^2 T_e / m_e \nu_{ei} \omega\right)\right]}{\left[1 + 5.1\left(i k_\|^2 T_e / m_e \nu_{ei} \omega\right) + 2.1\left(i k_\|^2 T_e / m_e \nu_{ei} \omega\right)^2\right]} u_{\|i},$$
(B.4)

where we have taken Z = 1 and

$$\sigma_0 = 1 - \frac{\omega_{*e}}{\omega}(1 + 1.71\eta_e), \qquad \sigma_1 = \left(1 - \frac{\omega_{*e}}{\omega}\right). \tag{B.5}$$

Equation (B.4) provides the semi-collisional Ohm's law:

$$j_\| = \frac{ne^2}{m_e \nu_{ei}} \frac{\left[\sigma_0 + d_1 \sigma_1 \left(i k_\|^2 T_e / m_e \nu_{ei} \omega\right)\right]}{\left[1 + d_0 \left(i k_\|^2 T_e / m_e \nu_{ei} \omega\right) + d_1 \left(i k_\|^2 T_e / m_e \nu_{ei} \omega\right)^2\right]} E_\|$$
$$- \frac{i k_\|^2 T_e}{m_e \nu_{ei} \omega} \frac{\left[d_2 + d_1 \left(i k_\|^2 T_e / m_e \nu_{ei} \omega\right)\right]}{\left[1 + d_0 \left(i k_\|^2 T_e / m_e \nu_{ei} \omega\right) + d_1 \left(i k_\|^2 T_e / m_e \nu_{ei} \omega\right)^2\right]} neu_{\|i},$$
(B.6)

with $d_0 = 5.1$, $d_1 = 2.1$ and $d_2 = 2.9$ and where the parallel electric field is given by

$$E_\| = i\omega A_\| - i k_\|(x)\varphi \ . \tag{B.7}$$

# Figures

**(a)**

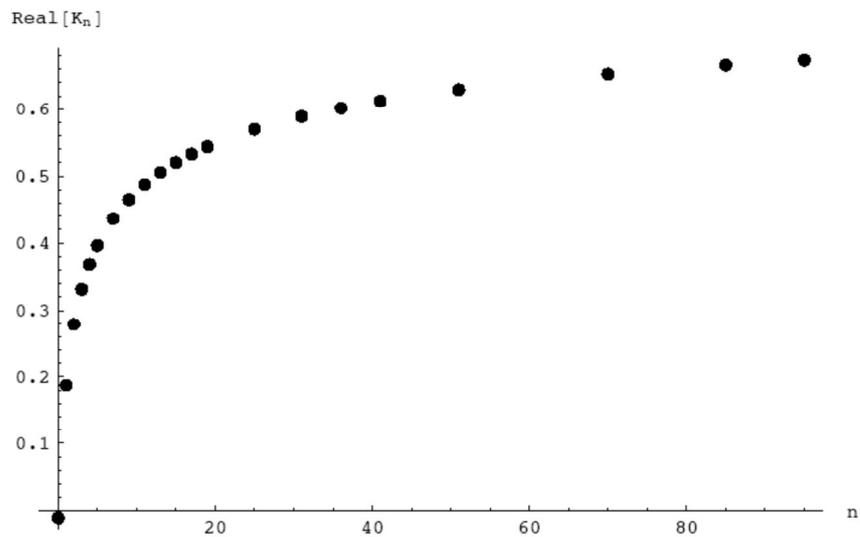

**(b)**

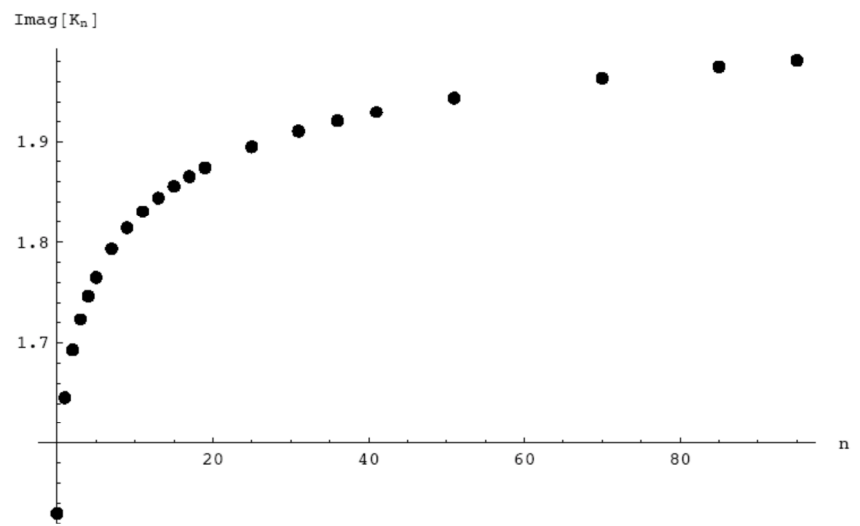

Fig.1. The integral $K_n$ as a function of n for the case $\alpha = \tau = 1$ : (a) the real part and (b) the imaginary part.

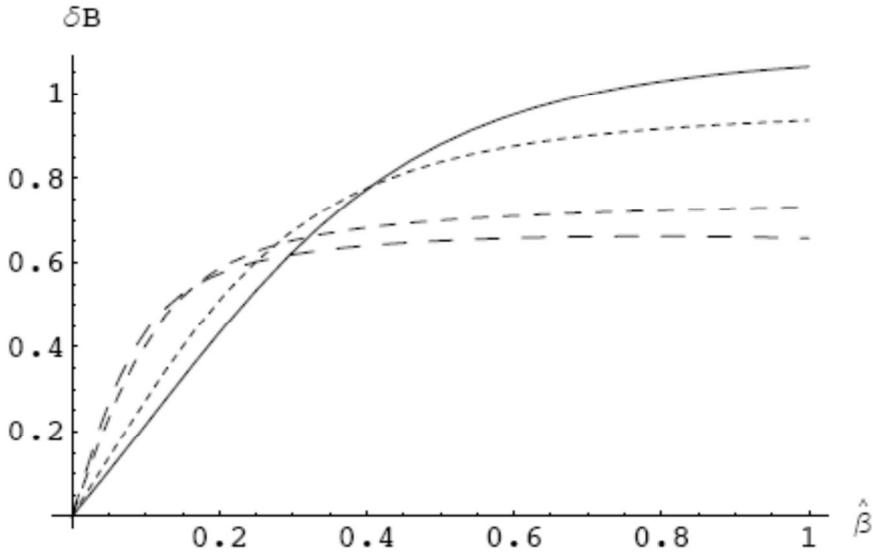

Fig.2. The variation of $\delta B_x(0)/B$, normalised to $e\varphi_0/T_i$ and in units of $2\times 10^{-5}$, with $\hat{\beta}_T$ for a number of n values: $\delta_0/\rho_i = 0.1$, $\alpha = \tau = 1$, $b_0 = 0.1$ and $L_{T_i} = L_{T_e}$. The solid line is n = 0 and the dashed lines correspond to n = 1, n = 11, n = 21, higher n being represented by longer dashes.

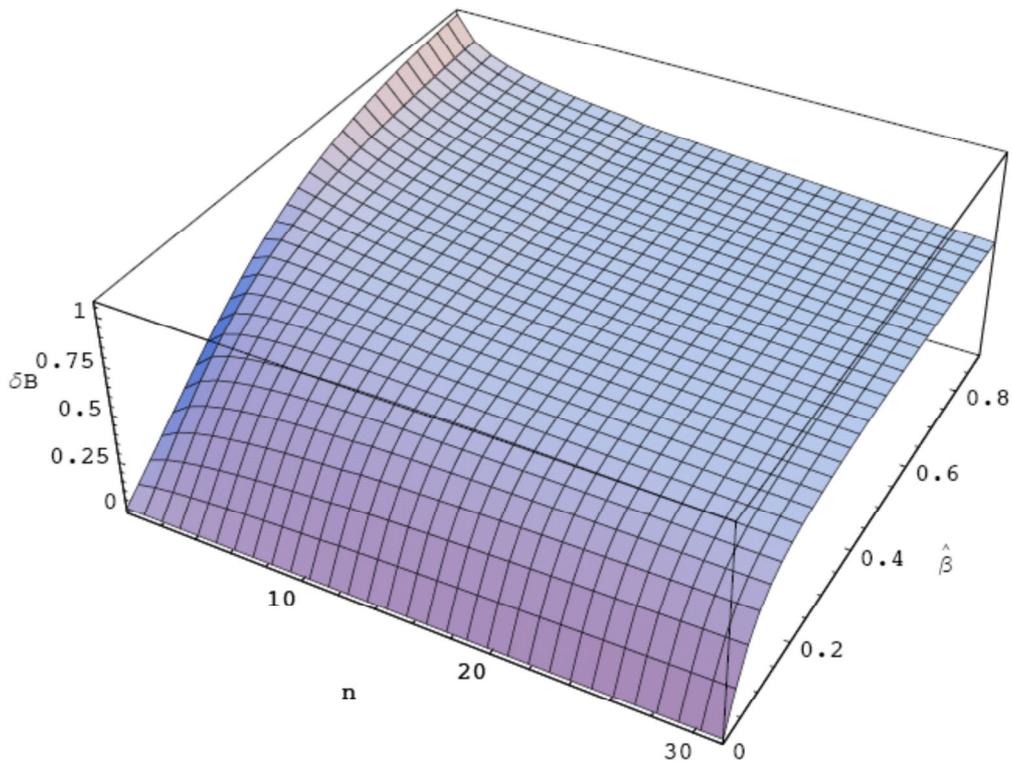

Fig.3. A three-dimensional plot showing the surface of $\delta B_x(0)/B$, normalised to $e\varphi_0/T_i$ and in units of $2\times 10^{-5}$, as a function of $\hat{\beta}_T$ and n for the case: $\delta_0/\rho_i = 0.1$, $\alpha = \tau = 1$, $b_0 = 0.1$ and $L_{T_i} = L_{T_e}$.

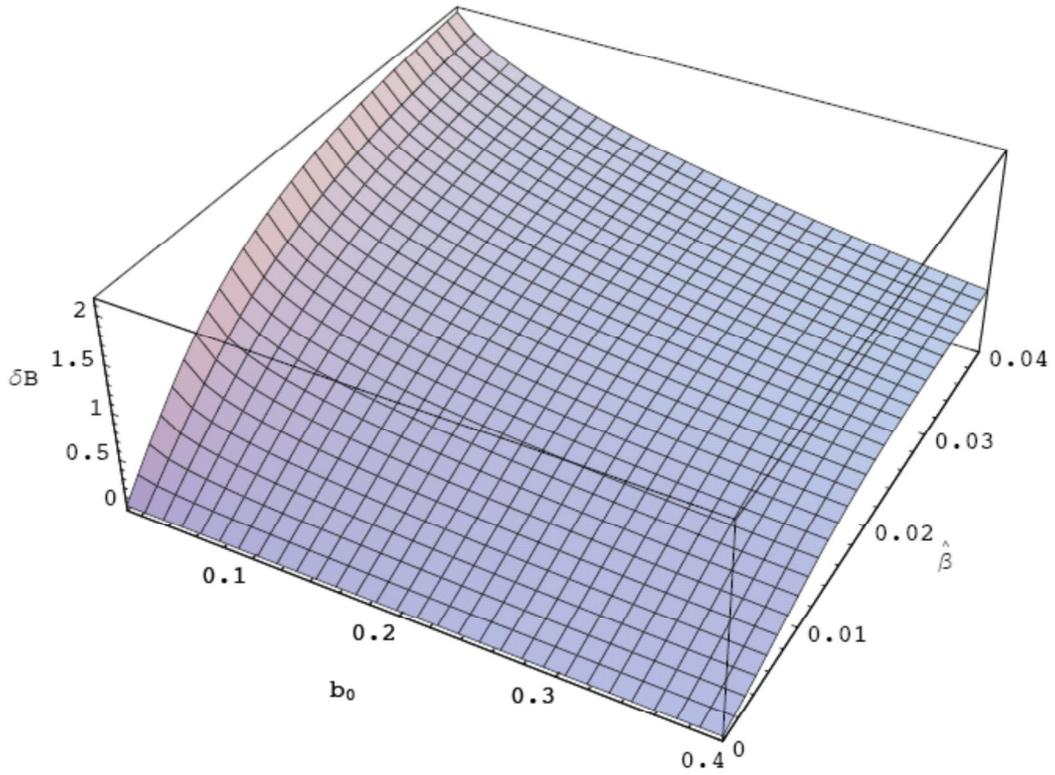

Fig.4. A three-dimensional plot showing the surface of $\delta B_x(0)/B$ normalised to $e\varphi_0/T_i$ and in units of $10^{-6}$, as a function of $\hat{\beta}_T$ and $b_0$ for the case: $n = 0$, $\hat{\delta} = 0.005$, $L_s/L_{Ti} = 25$, $\tau = 1$ and $L_{Ti} = L_{Te}$. The normalisation differs from Figs. 2 and 3 because of the smaller value of $\hat{\delta}$.